\documentclass[prd,aps,preprint,tightenlines,superscriptaddress,showpacs]{revtex4}
\usepackage{epsfig}
\usepackage{amsmath}

\newcommand{\beq}[1]{\begin{equation} \label{#1}}
\newcommand{\eeq}{\end{equation}}
\newcommand{\beqar}[1]{\begin{eqnarray} \label{#1}}
\newcommand{\eeqar}{\end{eqnarray}}
\newcommand{\beqs}{\begin{eqnarray}}
\newcommand{\eeqs}{\end{eqnarray}}
\newcommand{\bra}[1]{\langle {#1}|}
\newcommand{\ket}[1]{|{#1}\rangle}

\newcommand{\nid}{\noindent}
\newcommand{\non}{\nonumber}
\newcommand{\bib}{\bibitem}

\begin{document}

\title{Counting and Tensorial Properties of Twist-Two Helicity-Flip Nucleon Form Factors}

\author{Zhang Chen}
\email{chenz@mville.edu}
\affiliation{Department of Physics, Manhattanville College, Purchase, NY 10577, USA}
\author{Xiangdong Ji}
\email{xji@physics.umd.edu}
\affiliation{Department of Physics, University of Maryland, College Park, Maryland 20742, USA}
\date{\today}

\vspace{0.5in}

\begin{abstract}
We perform a systematic analysis on the off-forward matrix
elements of the twist-two quark and gluon helicity-flip operators.
By matching the allowed quantum numbers and their crossing channel
counterparts (a method developed by Ji \& Lebed), we
systematically count the number of independent nucleon form
factors in off-forward scattering of matrix elements of these
quark and gluon spin-flip operators. In particular, we find that
the numbers of independent nucleon form factors twist-2, helicity
flip quark (gluon) operators are $2n-1$ ($2n-5$) if $n$ is odd,
and $2n-2$ ($2n-6$) if $n$ is even, with $n\ge2$ ($n\ge 4$). We
also analysis and write down the tensorial/Lorentz structure and
kinematic factors of the expansion of these operators' matrix
elements in terms of the independent form factors. These
generalized form factors define the off-forward quark and gluon
helicity-flip distributions in the literature.
\end{abstract}



\maketitle

\section{Introduction}

To fully describe a nucleon in quantum chromodynamics (QCD), one
needs to know the matrix elements of all possible quark and gluon
operators involving the nucleon state. Among these operators, the
matrix elements of those of twist-two, often having clear physical
interpretation (e.g., corresponding to the energy-momentum
tensor), give the leading contribution (thus they are often
referred to as leading-twist) in appropriate hard processes. They
are also more accessible to experimental measurement and are
relatively simple.

The matrix elements can be taken between states of equal momenta
(forward) or unequal momenta (off-forward), and contain valuable
dynamical information about the internal structure of the nucleon.
In the forward case, after extracting out the tensorial/Lorentz
structure and kinematic factors, one obtains the irreducible
matrix elements. These are (combinations of) moments of, and thus
can be used to define, the conventional Feynman parton
distributions, (eg, see \cite{ref:Collins}) On the other hand, the
off-forward matrix elements are expanded in terms of (generalized)
nucleon form factors (see Eq. (\ref{eqn:qexpansion})), which are
closely connected to the moments of, and can be used to define a
new type of parton distributions--the (generalized) off-forward
parton distributions. At the same time, from the point view of the
low-energy nucleon structure, these off-forward distributions can
be considered as the generating functions for the form factors of
the twist-two operators. In recent years, these (generalized)
off-forward parton distributions, or simply generalized parton
distributions (GPDs), of hadrons, especially those of the nucleon,
have been the subject of much theoretical and experimental effort
\cite{ref:ofpd, ref:Diehl}.

As characterizations of certain properties exhibited by the
nucleon in classes of (most often non-forward) high-energy
scattering, e.g., deeply virtual Compton scattering (DVCS) and
diffractive electroproduction of vector mesons, GPDs represent the
low-energy internal structure of the particle. They are
generalizations of both the Feynman parton distributions and the
elastic electromagnetic form factors. In general the GPDs have
their physical interpretations in the Fourier space as the quantum
phase-space parton distributions \cite{ref:phasespace}. Most
interestingly, the distributions contain information about the
orbital motion of partons in a (polarized) nucleon. For instance,
knowing certain off-forward matrix elements and extracting the
related GPDs allows for deduction of the quark and gluon orbital
and spin contributions to the nucleon spin \cite{ref:spin}.

Therefore, the study of the generalized nucleon form factors is of
much interest and importance. One of the essential understandings
lies with the enumeration of independent nucleon form factors of
twist-two operators, as well as the Lorentz structure and
kinematic factors of the off-forward matrix elements, namely, its
expansion into form factors. In \cite{ref:count}, a method was
developed to systematically count the number of hadronic form
factors based on the partial wave formalism and crossing symmetry.
There the case for spin-independent operators was discussed. The
class of twist-two operators which depends on parton helicity
change are the subject of this paper. We will enumerate the number
of independent form factors for (the off-forward matrix elements
of) quark and gluon helicity-flip operators, and write down the
form factor expansion for the general quark operators.

The outline of the paper is as follows. In Section
\ref{sec:Definition}, we give a brief review of the definitions of
the twist-two operators and their matrix elements, as well as
their relationship to the GPDs. Section \ref{sec:qcount} contains
the systematic enumeration of independent form factors of the
quark operators, while Section \ref{sec:qexpansion} presents the
form factor expansion for the quark helicity-flip operators.
Section  \ref{sec:gexpansion} does the same for the gluon
operators. We conclude the paper by giving the summary and outlook
in Section \ref{sec:summary}. In Appendix \ref{sec:tensor} we
provide a general discussion on tensorial properties of these
operators and their representations. And in Appendix
\ref{sec:Time}, using the quark operators as examples, we give a
discussion on the possible constraints Hermiticity and time
reversal invariance requirements might impose.

\section{The Twist-Two Helicity-Flip Quark and Gluon Operators} \label{sec:Definition}

Using the now quite standard notation (see, eg, \cite{ref:ofpd},
\cite{ref:flip}), we parameterize the kinematics of the
off-forward scattering process as follows. The momenta and spins
of the initial and final nucleons are $P,S$ and $P',S'$,
respectively. The four-momentum transfer $\Delta^\mu=P'^\mu-P^\mu$
has both longitudinal and transverse components, and its invariant
is $t=\Delta^2$. Define a special system of coordinates in which
the average nucleon momentum $\overline{P}^\mu = (P'+P)^\mu/2$ is
collinear and in the $z$ direction. Further define, as usual, two
light-like four-vectors $p^\mu$ and $n^\mu$ with $p^2=n^2=0$ and
$p\cdot n=1$. We have
\begin{eqnarray}
\overline{P}^\mu &=& (P'+P)^\mu/2 \,=\, p^\mu + (\bar M^2/2) n^\mu\ , \nonumber \\
\Delta^\mu &=& P'^\mu-P^\mu \,=\,-2\xi(p^\mu-(\bar M^2/2) n^\mu)
+ \Delta_\perp^\mu\ ,\nonumber \\
\bar M^2 &=& M^2 -\Delta^2/4 \ .
\label{eqn:kinematics}
\end{eqnarray}
The initial nucleon and parton have longitudinal momentum fractions $1+\xi$
and $x+\xi$, respectively.

The following tower of twist-two operators is a generalization
of the electromagnetic current
\begin{equation}
    {\cal O}^{\mu_1\cdots\mu_n}_q =
       \overline{\psi}_q (0) i\stackrel{\leftrightarrow}{\cal D}^{(\mu_1}
        \cdots  i\stackrel{\leftrightarrow}{\cal D}^{\mu_{n-1}}
        \gamma^{\mu_n)} \psi_q (0)\ ,
\label{eqn:qsym}
\end{equation}
where all indices are symmetrized and traceless (indicated by
$(\cdots)$) and $\stackrel{\leftrightarrow}{\cal D} =
(\stackrel{\rightarrow}{\cal D} - \stackrel{\leftarrow}{\cal
D})/2. $ with ${\cal D}$ as the covariant derivative in QCD. An
expansion of the off-forward matrix elements of these operators
give rise to nucleon form factors whose linear combinations (with
powers of $\xi$ as coefficients) are the moments of the GPDs $H(x,
\xi, t)$ and $E(x, \xi,t)$. Similarly, there are five additional
towers of twist-two operators in QCD besides that in Eq. (\ref{eqn:qsym}):
\begin{eqnarray}
    \tilde {\cal O}^{\mu_1\cdots\mu_n}_q &= &
       \overline \psi_q i\stackrel{\leftrightarrow}{\cal D}^{(\mu_1}
        \cdots  i\stackrel{\leftrightarrow}{\cal D}^{\mu_{n-1}}
        \gamma^{\mu_n)} \gamma_5 \psi_q\ , \nonumber \\
    {\cal O}^{\mu_1\cdots\mu_n\alpha}_{qT} &= &
       \overline \psi_q i\stackrel{\leftrightarrow}{\cal D}^{(\mu_1}
        \cdots  i\stackrel{\leftrightarrow}{\cal D}^{\mu_{n-1}}
        \sigma^{\mu_n)\alpha} \psi_q\ , \nonumber \\
   {\cal O}^{\mu_1\cdots\mu_n}_g &= &
       F^{(\mu_1\alpha} i\stackrel{\leftrightarrow}{\cal D}^{\mu_2}
        \cdots  i\stackrel{\leftrightarrow}{\cal D}^{\mu_{n-1}}
        F_\alpha^{~\mu_n)} \ , \nonumber \\
  \tilde {\cal O}^{\mu_1\cdots\mu_n}_g &= &
       F^{(\mu_1\alpha} i\stackrel{\leftrightarrow}{\cal D}^{\mu_2}
        \cdots  i\stackrel{\leftrightarrow}{\cal D}^{\mu_{n-1}}
        i\tilde F_\alpha^{~\mu_n)} \ , \nonumber \\
    {\cal O}^{\mu_1\cdots\mu_n\alpha\beta}_{gT} &= &
       F^{(\mu_1\alpha} i\stackrel{\leftrightarrow}{\cal D}^{\mu_2}
        \cdots  i\stackrel{\leftrightarrow}{\cal D}^{\mu_{n-1}}
        F^{\mu_n)\beta}\ .
  \label{eqn:operators}
\end{eqnarray}
The corresponding GPDs are labelled by $\Big(\tilde H_{q}(x,\xi)$,
$\tilde E_{q}(x, \xi)\Big)$, $\Big(H_{Tq}(x,\xi)$, $E_{Tq}(x,
\xi)$, $\tilde H_{Tq}(x,\xi)$,
    $\tilde E_{Tq}(x, \xi) \Big)$,
$\Big( H_{g}(x,\xi)$, $E_{g}(x, \xi)\Big)$,
$\Big(\tilde H_{g}(x,\xi)$, $\tilde E_{g}(x, \xi)\Big)$,
and $\Big(H_{Tg}(x,\xi)$, $E_{Tg}(x, \xi)$, $\tilde H_{Tg}(x,\xi)$,
    $\tilde E_{Tg}(x, \xi) \Big)$,
respectively \cite{ref:ofpd, ref:Diehl}.

We concentrate our attention to the helicity-flip operators ${\cal
O}_{qT}$ and ${\cal O}_{gT}$ and their corresponding form factors
and GPDs. For example, it is known that for the lowest spin of
each kind, the GPDs arise from the following
definition\cite{ref:flip,ref:Diehl,ref:Diehl_Ori}
\begin{eqnarray}
    \int  \frac{d\lambda}{2\pi} e^{i\lambda x}
      \langle P'S'|\bar\psi_q(-\frac{1}{2}\lambda n)\sigma^{\mu \nu}
            \psi_q(\frac{1}{2}\lambda n)|PS \rangle
      &=& H_{Tq}(x,\xi)
        \overline{U}(P'S')\sigma^{\mu \nu} U(PS) \nonumber \\
    + \; \tilde H_{Tq}(x,\xi) \!\!\!\!&&\!\!\!\! \overline{U}(P'S')
        \frac{\overline{P}^{[\mu} i\Delta^{\nu ]}}{M^2} \,U(PS) \non \\
    + \; E_{Tq}(x,\xi) \!\!\!\!&&\!\!\!\! \overline{U}(P'S')
          \frac{\gamma^{[\mu}i\Delta^{\nu]}}{M} \, U(PS) \non \\
    + \; \tilde E_{Tq}(x,\xi) \!\!\!\!&&\!\!\!\! \overline{U}(P'S')
          {\gamma^{[\mu}i\overline{P}^{\nu]} \over M} \, U(PS) + ... \,
\label{eqn:lowest-q}
\end{eqnarray}
\begin{eqnarray}
    {1\over x}\int \frac{d\lambda}{2\pi} e^{i\lambda x}
        \langle P'S'|F^{(\mu\alpha}(-\frac{\lambda}{2}n)
       F^{\nu\beta)}(\frac{\lambda}{2}n)|PS\rangle
        &=& H_{Tg}(x,\xi) \overline{U}(P'S')\frac{\overline{P}^{([\mu} i\Delta^{\alpha ]}
            \sigma^{\nu\beta)}}{M} \,U(PS) \nonumber \\
    + \; \tilde H_{Tg}(x,\xi) \!\!\!\!&&\!\!\!\! \overline{U}(P'S')
        {\overline{P}^{([\mu} \Delta^{\alpha ]}\over M}
        {\overline{P}^{[\nu}\Delta^{\beta])} \over M^2} \, U(PS) \non \\
    + \; E_{Tg}(x,\xi) \!\!\!\!&&\!\!\!\! \overline{U}(P'S')
        {\overline{P}^{([\mu} \Delta^{\alpha ]}\over M}
        {\gamma^{[\nu}\Delta^{\beta])} \over M} \, U(PS) \non \\
    + \; \tilde E_{Tg}(x,\xi) \!\!\!\!&&\!\!\!\! \overline{U}(P'S')
        {\overline{P}^{([\mu} \Delta^{\alpha ]}\over M}
        {\gamma^{[\nu}\overline{P}^{\beta])} \over M} \, U(PS) + ... \,
\label{eqn:lowest-g}
\end{eqnarray}

\nid where the dependence of each distribution upon $t=\Delta^2$
and $Q^2$ is implicit. In the first equation, $[\mu\nu]$ denotes
anti-symmetrization of the two indices and the ellipses represent
higher twist structures. The quark helicity-flip distributions
$H_{Tq}$, $\tilde H_{Tq}$, $E_{Tq}$ and $\tilde E_{Tq}$ can be
selected by taking $\mu=+$ and $\nu=\perp$. The gauge link between
the quark fields is not explicitly shown. Also by time-reversal
symmetry and Hermiticity (complex conjugate), the quark
distributions are real and even functions of $\xi$. In the second
equation $[\mu\alpha]$ and $[\nu\beta]$ are antisymmetric pairs
and $(\cdots)$ signifies symmetrization of the two and removal of
the trace. Similarly, the gluon helicity-flip distributions
$H_{Tg}$, $\tilde H_{Tg}$, $E_{Tg}$ and $\tilde E_{Tg}$ can be
selected by taking $\mu=\nu=+$ and $\alpha,\beta = \perp$.

\section{Counting of Independent Form Factors of Quark Operators}\label{sec:qcount}

The quark and gluon operators defined above transform as
irreducible representations of the Lorentz group. That is the
reason for the (anti)symmetrization of the indices and removal of
the traces. In this section we first briefly discuss the
enumeration of independent elements of tensors with certain
symmetry type and trace conditions. Then we consider the number of
independent form factors for the matrix elements of the
corresponding operators.

The number of independent elements of a tensor of rank $n$ in
$n$-dimensional (or $d$-dim) space is $d^n$. 
That of a totally symmetric tensor is $C^{d+n-1}_n
=(d+n-1)!/(n!(d-1)!)$,
and that of a totally antisymmetric tensor is $C^d_n
=d!/(n!(d-n)!)$.
The number of traceless conditions for a generic rank $n$ tensor
($n \ge 2$) (in $d$-dim) is $C^n_2d^{n-2}$,
while that for a rank $r$ symmetric tensor is $C^{d+n-3}_{n-2}$.
(This can be easily shown from recognizing that this number is the
same as the number of independent elements of a rank $n-2$
symmetric tensor, since a traceless condition is simply
contracting two indices of the tensor.) Therefore, the number of
independent elements of a rank-$n$ symmetric traceless tensor in
$n$-dim is $ C^{d+n-1}_r - C^{d+n-3}_{n-2}= (n+1)^2$. In the
notation of \cite{ref:Hamermesh}, where square parentheses are
used to denote (the number of independent elements of) tensors
with traces and usual (round) parentheses are used to denote (that
of) traceless tensors, and the first number in a parenthesis is
the number of symmetrized indices and the second number is that of
anti-symmetrized indices, the above conclusion can be written as
$$(n,0) = [n,0] - [n-2, 0] \;.$$
The same result can be obtained by recognizing that such tensors
furnish $\{\frac{n}{2}, \frac{n}{2}\}$ representations of the
Lorentz group \cite{ref:Landau, ref:Weinberg}, with their elements
written as $T^{\alpha_1 \cdots \;
\alpha_n}_{\;\;\;\;\;\;\;\;\;\;\;\; \dot{\beta}_1 \cdots \;
\dot{\beta}_n}$ (here we use curly brackets instead of usual
parentheses to avoid confusion with the above notation). Such
representations $\{A,B\}$ have $(2A+1)(2B+1)$ independent
elements. As far as the $J^{PC}$ properties and the number of
independent matrix elements of the corresponding operators are
concerned, they have one to one correspondence with Weyl spinors
$\{A,B\}$ (e.g., \cite{ref:Landau, ref:Weinberg}).

For example, the symmetric quark operators in Eq. (\ref{eqn:qsym}),
\beq{eqn:symop}
    {\cal O}^{\mu_1 ... \mu_n}_{q} = \overline{\psi}(0)
    i \stackrel {\leftrightarrow}{\cal D}^{(\mu_1} \cdots \;
    i \stackrel{\leftrightarrow}{\cal D}^{\mu_{n\!-\!1}}
    \gamma^{\mu_n)} \psi(0)\ ,
\eeq
\nid form totally symmetric tensor representations of the
Lorentz group. Each ${\cal O}_{qT}^{\mu_1\cdots\mu_n}$ is a tensor
of $(n,0)$, and furnishes a $\{n/2,n/2\}$ representation of
Lorentz group and has $(n+1)^2$ independent components.\cite{ref:Weinberg}

On the other hand, the operators ${\cal O}_{qT}$ in
(\ref{eqn:operators}) are rank $d=n+1$ traceless tensors in 4-dim
with symmetric indices $\mu_1, ..., \mu_n$ and are antisymmetric
in $\mu_n, \alpha$. For such operators
\beq{eqn:qasymop}
    {\cal O}^{\mu_1 ... \mu_{n-1} \mu_n \alpha}_{qT} = \overline{\psi}(0)
        i \stackrel {\leftrightarrow}{\cal D}^{(\mu_1} \cdots \;
        i \stackrel{\leftrightarrow}{\cal D}^{\mu_{n-1}} \sigma^{\mu_n) \alpha} \psi(0),
\eeq
\nid the number of independent elements, labelled $(n,1)$,
can be worked out by the tensor method as shown in Appendix
\ref{sec:tensor}. The
result is $(n,1)=2\times n(n+2)$. This is the same as the number
of independent elements of the representation $\{\frac{n-1}{2}$,
$\frac{n+1}{2}\}$ (with element $T^{\alpha_1 \cdots \;
\alpha_{n\!-\!1}}_{\;\;\;\;\;\;\;\;\;\;\;\;\;\;\; \dot{\beta}_1
\cdots \;\dot{\beta}_{n\!+\!1}}$) and (plus) $\{\frac{n+1}{2}$,
$\frac{n-1}{2}\}$ (with element $T^{\alpha_1 \cdots \;
\alpha_{n\!+\!1}}_{\;\;\;\;\;\;\;\;\;\;\;\;\;\;\; \dot{\beta}_1
\cdots \; \dot{\beta}_{n\!-\!1}}$).

The $J^{PC}$ content of the operators (\ref{eqn:qsym}) and the
numeration of the resulting independent form factors were
discussed in \cite{ref:count}. Following the same method, one can
analyze the operators in Eq. (\ref{eqn:qasymop}).

First one goes to the crossed channel where the operator serves as a source
for creating a particle-antiparticle pair (matrix elements
$\bra{P\bar{P}} {\cal O}^{\mu_1 ... \mu_k \mu \alpha} \ket{0}$, where $k+1 \equiv n$).
The allowed $J^{PC}$ values in this case are enumerated as the following
(similar to the
discussion in \cite{ref:count}): For $J^{PC}(L)$ values of
$P\bar{P}$, $P=(-1)^{L+1}$, $C=(-1)^{L+S}$,
and $S=0, 1$. Thus in terms of $J$, when
$S=0$: $L=J$, $P=(-1)^J+1$, $C=(-1)^J, (-1)^{J+1}$,
and when $S=1$: $L=J\!\pm\!1$, $P=(-1)^J$, $C=(-1)^J$.


The $J^{PC}$ of the operators (\ref{eqn:qasymop}) can be classified
as the following: The representation $\{A, B\}$ has angular
momentum $J = |A-B|, |A-B|+1, ..., A+B$ (since $\vec{J} = \vec{A}
+ \vec{B}$). The natural parity of the operator is $P=(-1)^J$,
while the charge conjugation $C$ of $\gamma^\mu$, (each) $i
\stackrel{\leftrightarrow}{\cal D}^\mu$, and $\sigma^{\mu\nu}$ are
all $-1$. Since parity transforms $A \leftrightarrow B$, for each
$J$, both $\pm$ parities are allowed.
Then the representations $\{\frac{n-1}{2}$, $\frac{n+1}{2}\}$ and
$\{\frac{n+1}{2}$, $\frac{n-1}{2}\}$ have $J = 1, 2, ...\;, n$,
$P=\pm$ and $C = (-1)^n$. The $J^{PC}$ content of the operators
and the the $J^{PC}(L)$ values of the cross channel $P\bar{P}$
system are

\begin{tabbing}
Social Sec \= Social Security aaaaaaaaaaaaaaaaaaaaaaaaaaa \= \kill
$n$ \> $\;\;\;\;\;\;{\cal O}^{\mu_1...\mu_n\nu}$
    \> $\;\;\;\;\;\; P\bar{P} \; (J^{PC}(L))$ \\
\>\\
$\;\; 1$ \> $1^{+-}, ~1^{--}$ \> $1^{++}(1), ~1^{+-}(1), ~1^{--}(0), ~1^{--}(2)$ \\
$\;\; 2$ \> $1^{++}, ~1^{-+}, ~2^{++}, ~2^{-+}$
    \> $2^{++}(1), ~2^{++}(3), ~2^{-+}(2), ~2^{--}(2)$\\
$\;\; 3$ \> $1^{+-}, ~1^{--}, ~2^{+-}, ~2^{--}, ~3^{+-}, ~3^{--}$
    \> $3^{++}(3), ~3^{+-}(3), ~3^{--}(2), ~3^{--}(4)$\\
$\;\; 4$ \> $1^{++}, ~1^{-+}, ~2^{++}, ~2^{-+}, ~3^{++}, ~3^{-+}, ~4^{++}, ~4^{-+}$
    \> $4^{++}(3), ~4^{++}(5), ~4^{-+}(4), ~4^{--}(4)$\\
. \> ... \> ... \\
$\;\; n$ \> $1^{+(-)^n}, ~1^{-(-)^n}, ~... \; , ~n^{+(-)^n}, ~n^{-(-)^n}$
    \> $n^{(-)^n(-)^n}(n\!-\!1), ~n^{(-)^n(-)^n}(n\!+\!1),$ \\
\>  \> $n^{(-)^{n+1}(-)^n}(n), ~~n^{(-)^{n+1}(-)^{n+1}}(n)$\\
i.e. \\
$n = odd$ \> $1^{+-}, ~1^{--}, ~... \;, ~n^{+-}, ~n^{--}$
    \> $n^{--}(n\!\pm\!1), ~n^{++}(n), ~n^{+-}(n)$\\
$n = even$ \> $1^{++}, ~1^{-+}, ~... \;, ~n^{++}, ~n^{-+}$
    \> $n^{++}(n\!\pm\!1), ~n^{-+}(n), ~n^{--}(n)$\\
\end{tabbing}

For each $J^{PC}$, the number of independent form factors in
matrix elements $\bra{P'}{\cal O}^{\mu_1 ... \mu_n \alpha}\ket{P}$ is determined
by the number of independent amplitudes for the creation process (in the
cross channel). For example, for $n=1$, both $1^{+-}$ and $1^{--}$
sources are effective. While $1^{+-}$ can only create one state, the
$1^{--}$ source can create two independent states (with $L=0$ and
$2$). Therefore, there are three independent matrix elements.
For $n=2$, sources $1^{++}$, $2^{++}$, and $2^{-+}$ are effective.
Both $1^{++}$ and $2^{-+}$ sources can only create one state, while
the $2^{++}$ source can again produce two states (with $L=1$ and $3$).
So there are four independent matrix elements. A list of matrix elements
in terms of these quantum numbers can be generated as
\begin{eqnarray}
&& n=1, ~J^{PC}(L) = 1^{+-}(1), ~1^{--}(0), ~1^{--}(2)   \nonumber  \\
&& n=2, ~J^{PC}(L) = 1^{++}(1); ~2^{++}(1), ~2^{++}(3), ~2^{-+}(2)   \nonumber \\
&& n=3, ~J^{PC}(L) = 1^{+-}(1), ~1^{--}(0), ~1^{--}(2); ~2^{--}(2);
    ~3^{+-}(3), ~3^{--}(2), ~3^{--}(4) \nonumber \\
&& n=4, ~J^{PC}(L) = 1^{++}(1); ~2^{++}(1), ~2^{++}(3), ~2^{-+}(2);
    ~3^{++}(3); ~4^{++}(3), ~4^{++}(5), ~4^{-+}(4) \nonumber\\
&& \cdots . \nonumber
\end{eqnarray}

\nid This pattern can be extended and we have the enumeration of the
independent form factors in $\bra{P'}{\cal O}^{\mu_1 ... \mu_k \mu \nu}\ket{P}$
as, where the $(\times 2)$ represents the two different $L$ ($=J\pm1$) values for
each $J$,

\begin{tabbing}
Social Sec \= Social Security aaaaaaaaaaaaaaaaaaaaaaaaaaa \= \kill
$n$ \> $\;\;\;\;\;\;{\rm Matched} \; J^{PC}$
    \> $\;\;\;\;\;\; {\rm Total} \;\; {\rm Number}$ \\
\>\\
$\;\; 1$ \> $1^{+-}, ~1^{--} (\times 2) $ \> $ (1+2) = 3 $ \\
$\;\; 2$ \> $1^{++}; ~2^{++} (\times 2), ~2^{-+}$ \> $(1) + (1+2) = 4$\\
$\;\; 3$ \> $1^{+-}, ~1^{--} (\times 2); ~2^{--}; ~3^{+-}, ~3^{--} (\times 2)$
    \> $ (1+2) + (1) + (1+2) = 7$\\
$\;\; 4$ \> $1^{++}; ~2^{++} (\times 2), ~2^{-+}; ~3^{++}; ~4^{++} (\times 2), ~4^{-+}$
    \> $(1) + (1+2) + (1) + (1+2) = 8$\\
\>\\
$\;\;$... \> ... \> ... \\
\>\\
$n = odd$ \> $1^{+-}, ~1^{--} (\times 2); ~2^{--}; ~... \;; ~n^{+-}, ~n^{--} (\times 2)$
    \> $ (1+2)+(1)+ ... +(1+2) $ \\
    \> \> $= \frac{n+1}{2} \times 3 + \frac{n-1}{2} = 2n\!+\!1 $\\
\>\\
$n = even$ \> $1^{++}; ~2^{++} (\times 2), ~2^{-+}; ~... \;; ~n^{++} (\times 2), ~n^{-+}$
    \> $ (1) + (1+2) + ... +(1+2) $ \\
    \> \> $= \frac{n}{2} \times 3 + \frac{n}{2} = 2n $\\
\end{tabbing}

Time reversal invariance does not impose any constraint in the
crossed channel counting. However, when the same result is applied
to the direct channel, the number reflects that after applying the
time reversal symmetry.
(Please also see discussions in Section \ref{sec:qexpansion}
and Appendix \ref{sec:Time}.)

\section{Form Factors of Twist-Two Helicity-Flip Quark Operators} \label{sec:qexpansion}

As was indicated in \cite{ref:count,ref:Diehl}, time reversal
invariance might (or might not) introduce new constraints on the
form factors thus further limit their numbers. In appendix
\ref{sec:Time} we look at the effect it has on the helicity-flip
operators, and explicitly showed that it does not pose further
limits on the matrix elements of the lowest rank operator in Eq.
(\ref{eqn:qasymop}), and there are only four types of terms in the
expansion of the matrix elements (see for example, Eq.
(\ref{eqn:OpHermf})). The higher rank operators in
(\ref{eqn:qasymop}) (with additional covariant derivatives) will
have factors of $\overline{P}$ and $\Delta$ after taking matrix
elements between $p'$ and $p$, coming from the covariant
derivatives\cite{ref:ofpd, ref:CZ}. From Hermiticity requirements,
after taking the hermitian/time reversal, each factor of
$\Delta^\mu$ will introduce a factor of $-1$ while $p^\mu$ factor
will not. Therefore, the overall sign factor resulting from the
time reversal/Hermitian operation that comes from the covariant
derivatives (plus the extra factor of $\Delta$ from the
anti-symmetric part in the $C_3$ and $C_4$ terms), in the matrix
elements, will be $(-1)^l$ with $l$ the number of factors of
$\Delta$ present.

For the matrix element
$\bra{P'} {\cal O}^{\alpha \mu_1 \mu_2 \cdots \mu_n} \ket{P}$,
(we have made a rearrangement of the indices that result only in a
possible overall negative sign), just like the discussion leading to
(\ref{eqn:oexp}), there are still only four types of terms in the
expansion of the matrix element because of their Lorentz/tensorial
structure, namely terms of the format
$\overline{U} [\gamma^\alpha, \gamma^{\mu_1}] U \sim \sigma^{\mu\nu}$
    (similar to the $C_1$ term in (\ref{eqn:OpHermf})),
$\overline{U} [\gamma^\alpha, \overline{P}^{\mu_1}] U \sim \gamma^\mu$
    (similar to the $C_2$ term in (\ref{eqn:OpHermf})),
$\overline{U} [\gamma^\alpha, \Delta^{\mu_1}] U \sim \gamma^\mu$
    (similar to the $C_3$ term in (\ref{eqn:OpHermf})), and
$\overline{U} [\overline{P}^\alpha, \Delta^{\mu_1}] U \sim \overline{U}U$
    (similar to the $C_4$ term in (\ref{eqn:OpHermf})).
The $\sim$ sign means having the same properties under time
reversal (hermitian, complex conjugate). The operator ${\cal
O}^{\alpha \mu_1 \mu_2 \cdots \mu_n}$ is odd (``$-$") overall
under time reversal because of the anti-symmetric part
$\sigma_{\mu\nu}$ (the covariant derivatives are even). Same is
the first type of terms, while the rest three are even (``+").
Therefore to give the matrix elements the proper signs under time
reversal/Hermitian, namely overall odd (``$-$"), only certain
numbers of factors of $\Delta$ would appear thus limiting the
numbers of form factors and the structures associated with them.
We thus have the following table:

\begin{table}[ht]
\vspace{6pt} \caption{Enumeration from time-reversal/Hermiticity
considerations} \label{tab:hermiticity}
\vspace{12pt}
\begin{center}
\begin{tabular}{||c|c|c|c|c|c||}

\hline \hline
    Term & $~\hat{\mathcal{T}}~$
    & \multicolumn{4}{|c||}{Sign allowed from Factors of $\overline{P}$ and $\Delta$} \\
    \cline{3-6}
& & $~n\!=\!1~$ & $\cdots$ & $ \;\;\; n=2k+1$ & $ \;\;\; n=2k$ \\
\hline \hline
$\overline{U} [\gamma^\alpha, \gamma^{\mu_1}] U$ & $ - $ & $
\;\; + $
    && $  \;\; (-1)^l,$ with & $ \;\; (-1)^l,$ with \\
&&&& $~l=0, 2, \cdots, 2k (= n\!-\!1)$ & $~l0, 2, \cdots, 2k\!-\!2 (= n\!-\!2)$ \\
\hline
\# allowed &&$\;\; 1$&& $  \;\;\; k+1 = \frac{n+1}{2}$ &$ \;\; k = \frac{n}{2}$ \\
\hline \hline
$\overline{U} [\gamma^\alpha, \overline{P}^{\mu_1}] U$ & $ + $ &
$ \;\; + $
    && $  \;\; (-1)^l,$ with & $ \;\; (-1)^l,$ with\\
&&&& $~l = 1, 3, \cdots, 2k\!-\!1 (= n\!-\!2)$ & $~l = 1, 3, \cdots, 2k\!-\!1 (= n\!-\!1)$ \\
\hline
\# allowed &&$\;\; 0$&& $  \;\;\; k = \frac{n-1}{2}$ &$ \;\; k = \frac{n}{2}$ \\
\hline\hline
$\overline{U} [\gamma^\alpha, \Delta^{\mu_1}] U$ & $ + $ & $
\;\; - $
    && $  \;\; (-1)^l,$ with & $ \;\; (-1)^l,$ with\\
&&&& $~l = 0, 2, \cdots, 2k (= n\!-\!1)$ & $~l = 0, 2, \cdots, 2k\!-\!2 (= n\!-\!2)$ \\
\hline
\# allowed &&$\;\; 1$&& $  \;\;\; k+1 = \frac{n+1}{2}$ &$ \;\; k = \frac{n}{2}$ \\
\hline\hline
$\overline{U} [\overline{P}^\alpha, \Delta^{\mu_1}] U$ & $ + $ &
$ \;\; - $
    && $  \;\; (-1)^l,$ with & $ \;\; (-1)^l,$ with\\
&&&& $~l = 0, 2, \cdots, 2k (= n\!-\!1)$ & $~l = 0, 2, \cdots, 2k\!-\!2 (= n\!-\!2)$ \\
\hline
\# allowed &&$\;\; 1$&& $  \;\;\; k+1 = \frac{n+1}{2}$ &$ \;\; k = \frac{n}{2}$ \\
\hline \hline
Total \# && $3$ && $\; 4k\!+\!3 = 2n\!+\!1$ & $ \; 4k = 2n$ \\
\hline\hline
\end{tabular}
\end{center}
\end{table}

\newpage

The above is completely consistent and thus confirms our counting of the numbers of independent
form factors of the quark operators in section \ref{sec:qcount}. Further, it is now clear that
the matrix elements of these operators must have its expansion into kinematic factors,
Lorentz/tensorial (symmetry) structure factors and independent form factors in the form of
\beqar{eqn:qexpansion}
    \bra{P'} {\cal O}^{\alpha \mu_1 \mu_2 \cdots \mu_n} \ket{P}
        &=& \overline{U}(P')\;\sigma^{\alpha (\mu_1} \; U(P)
                \; \sum_{i=0}^{[\frac{n+1}{2}]}
            A_{n,2i\!-\!1} \; \Delta^{\mu_2}\Delta^{\mu_3} \cdots \Delta^{\mu_{2i\!-\!1}}
                \overline{P}^{\mu_{2i}} \cdots \overline{P}^{\mu_n )} \non \\
        && + \; \overline{U}(P') \; [\gamma^\alpha, \overline{P}^{(\mu_1}] \; U(P)
                \; \sum_{i=0}^{[\frac{n}{2}]}
            B_{n,2i} \; \Delta^{\mu_2}\Delta^{\mu_3} \cdots \Delta^{\mu_{2i}}
                \overline{P}^{\mu_{2i\!+\!1}} \cdots \overline{P}^{\mu_n)} \non \\
        && + \; \overline{U}(P') \; [\gamma^\alpha, \Delta^{(\mu_1}] \; U(P)
                \; \sum_{i=0}^{[\frac{n+1}{2}]}
            i C_{n,2i\!-\!1} \; \Delta^{\mu_2}\Delta^{\mu_3} \cdots \Delta^{\mu_{2i\!-\!1}}
                \overline{P}^{\mu_{2i}} \cdots \overline{P}^{\mu_n)} \non \\
        && + \; \overline{U}(P') \; [\overline{P}^\alpha, \Delta^{(\mu_1}] \; U(P)
                \; \sum_{i=0}^{[\frac{n+1}{2}]}
            i D_{n,2i\!-\!1} \; \Delta^{\mu_2}\Delta^{\mu_3} \cdots \Delta^{\mu_{2i\!-\!1}}
                \overline{P}^{\mu_{2i}} \cdots \overline{P}^{\mu_n)} \;. \non \\
\eeqar One can recover Eq. (\ref{eqn:lowest-q}) easily by multiplying the above by
$n^{\mu_1}n^{\mu_2}...n^{\mu_{n-1}}$ and converting the moments
into the light-cone fraction space.

\section{Form Factors of Twist-Two Helicity-Flip Gluon Operators} \label{sec:gexpansion}

The tensor operators that flips gluon spin are
\beq{eqn:goperator}
    {\cal O}^{\mu \alpha \nu \beta \mu_1 ... \mu_n}_{gT} = F^{(\mu \alpha}
        i \stackrel {\leftrightarrow}{\cal D}^{\mu_1} \cdots \;
        i \stackrel{\leftrightarrow}{\cal D}^{\mu_n} F^{\nu) \beta},
\eeq \nid where $\mu, \nu$ and $\mu_1, ..., \mu_n$ are symmetrized
while $\mu$ and $\alpha$ as well as $\nu$ and $\beta$ are
anti-symmetric pairs, and the operator is also rendered traceless
(note we relabelled $ {\cal O}_{gT}$ in (\ref{eqn:operators})
using $\mu_1 \rightarrow \mu$, $\mu_n \rightarrow \nu$, and
{$\mu_2, \cdots, \mu_{n\!-\!1}$} to {$\mu_1, \cdots, \mu_n$}).
Equation (\ref{eqn:goperator}) corresponds to the Young tableau
$(n+2, 2)$, as well as the Weyl representations $\{\frac{n}{2},
\frac{n+4}{2}\}$ (with element labelled $T^{\alpha_1 \cdots \;
\alpha_n}_{\;\;\;\;\;\;\;\;\;\;\;\; \dot{\beta}_1 \cdots
\;\dot{\beta}_{n\!+\!4}}$) and $\{\frac{n+4}{2}, \frac{n}{2}\}$ (
with elements labelled $T^{\alpha_1 \cdots \;
\alpha_{n\!+\!4}}_{\;\;\;\;\;\;\;\;\;\;\;\;\;\;\;\; \dot{\beta}_1
\cdots \;\dot{\beta}_n}$). That is,

\vspace*{0.3cm}

\unitlength0.5cm
\begin{picture}(30,1)
\linethickness{0.075mm}
\put(1,-1){\framebox(1,1){$\alpha$}}
\put(1,0){\framebox(1,1){$\mu$}}
\put(2,0){\framebox(1,1){$\nu$}}
\put(2,-1){\framebox(1,1){$\beta$}}
\put(3,0){\framebox(1,1){$\mu_1$}}
\put(4,0){\framebox(3,1){$\ldots$}}
\put(7,0){\framebox(2,1){$\mu_n$}}
\put(9.5,0){$- \; \mathrm{trace~term}$} \put(14.5,0){$=$}
\put(15.5,0){$(n\!+\!2,2)$ \;.}
\end{picture}

\vspace*{1cm}

According to the standard group theoretical result,
$[n+2,2]=(n+1)(n+4)(n+5)$. As shown in Appendix A, the number of
trace conditions for the tensor $[n+2,2]$ is $(n+1)(n+2)(n+5)$,
resulting in the number of independent elements of the traceless
tensor $(n+2,2)$ being $2(n+1)(n+5)$. This is verified by the
enumeration of the elements of the Weyl representations
$\{\frac{n}{2}, \frac{n+4}{2}\}$ and $\{\frac{n+4}{2},
\frac{n}{2}\}$, each of which being $(n+1)(n+5)$.

Now let us discuss the $J^{PC}$ content of these gluon
helicity-flip operators. The angular momentum $J$ obviously can
take on values of $2, 3, \cdots, n\!+\!2$. Similar to the
discussion in section \ref{sec:qcount}, both parity values are
allowed. The charge conjugation of the gluon field bilinear $FF$
is even ("+") (while $F^{\mu\nu}$ transforms just like
$\sigma^{\mu\nu}$ under time reversal \cite{ref:IZ, ref:Peskin}).
Therefore comparing with the allowed $J^{PC}$ values in the cross
channel, just as was done in section \ref{sec:qcount}, we have the
following enumeration for the number of independent form factors,

\begin{tabbing}
Social Sec \= Social Security aaaaaaaa \= Social Security aaaaaaaaaaaaaaaa \= \kill
$\;n$ \> $\;\;\;\;\; {\cal O}^{\mu \alpha \nu \beta \mu_1 ... \mu_n}$
    \> $\;\;\;\;\;\; {\rm Matched} \; (J^{PC}(L))$ \> ${\rm Enumeration}$  \\
\>\\
$\;\; 0$ \> $2^{++}, ~2^{-+}$
    \> $2^{++}(1), ~2^{++}(3), ~2^{-+}(2)$ \> $ \;\;\;\;\;\; 3$\\
$\;\; 1$ \> $2^{+-}, ~2^{--}, ~3^{+-}, ~3^{--}$
    \> $2^{--}(2), ~3^{+-}(3), ~3^{--}(2), ~3^{--}(4)$ \> $1+3 =4$\\
$\;\; 2$ \> $2^{++}, ~2^{-+}, ~3^{++}, ~3^{-+}$
    \> $2^{++}(1), ~2^{++}(3), ~2^{-+}(2), ~3^{++}(3)$ \> $3+1+3$\\
        \> $4^{++}, ~4^{-+}$
    \> $4^{++}(3), ~4^{++}(5), ~4^{-+}(4)$ \> $\;\;\;\; = 7$\\
$\;\; 3$ \> $2^{+-}, ~2^{--}, ~3^{+-}, ~3^{--}$
    \> $2^{--}(2), ~3^{+-}(3), ~3^{--}(2), ~3^{--}(4)$ \> $1+3+1$ \\
        \> $4^{+-}, ~4^{--}, ~5^{+-}, ~5^{--}$
    \> $4^{--}(4), ~5^{+-}(5), ~5^{--}(4), ~5^{--}(6)$ \> $\; +3 = 8$ \\
$\cdots$ \> $\cdots$ \> $\cdots$ \> $\cdots$  \\
\>\\
$n = {\rm odd}$ \> $2^{+-}, ~2^{--}, ~... \;,$ \> $ 2^{--}(2),
~3^{+-}(3), ~3^{--}\!\times\!2, \cdots,$
        \> $(1+3)\times \frac{n\!+\!1}{2} $\\
    \> $[n\!+\!2]^{+-}, ~[n\!+\!2]^{--}$
        \> $[n\!+\!2]^{+-}(n\!+\!2), ~[n\!+\!2]^{--} \times 2$ \> $=2(n\!+\!1)$\\
\>\\
$n = {\rm even}$ \> $2^{++}, ~2^{-+}, ~... \;,$ \> $2^{++} \times
2, ~2^{-+}(2), \cdots,$
        \> $ 3+(1+3)\!\times\!\frac{n}{2}$\\
    \> $[n\!+\!2]^{++}, ~[n\!+\!2]^{-+}$ \> $[n\!+\!2]^{++} \times 2, ~[n\!+\!2]^{-+}(n\!+\!2)$
        \> $ = 2n\!+\!3 $\\
\end{tabbing}
We notice that the number of form factors for the gluon operators
is the same as that for the quark operators with corresponding
spin.

It is obvious from comparison of Eq. \ref{eqn:lowest-q} and Eq.
\ref{eqn:lowest-g} that the only difference between the kinematic
and tensorial factors of the quark and gluon matrix elements is a
factor of $\overline{P} i \Delta$. More detailed analysis confirms
this, and following the similar discussions in sections
\ref{sec:qcount},\ref{sec:qexpansion} and appendix \ref{sec:Time},
we find that the form factor expansion of the matrix elements of
the twist-two helicity-flip gluon operators is
\beqar{eqn:gexpansion}
    \bra{P'} {\cal O}^{\mu \alpha \nu \beta \mu_1 ... \mu_n}_{gT} \ket{P}
        &=& \overline{U}(P') \;\sigma^{(\mu\alpha} \; U(P)
                \; \sum_{i=0}^{[\frac{n}{2}]}
            i A_{n,2i\!-\!1} \; \Delta^{\mu_1}\Delta^{\mu_2} \cdots \Delta^{\mu_{2i\!-\!1}}
                \overline{P}^{\mu_{2i}} \cdots \overline{P}^{\mu_n}
                \; [\overline{P}^{\nu)}, \Delta^{\beta }] \non \\
        + && \!\!\!\!\!\!
            \overline{U}(P') \; [\gamma^{(\mu}, \overline{P}^{\alpha}] \; U(P)
                \; \sum_{i=0}^{[\frac{n-1}{2}]}
            i B_{n,2i} \; \Delta^{\mu_1}\Delta^{\mu_2} \cdots \Delta^{\mu_{2i}}
                \overline{P}^{\mu_{2i\!+\!1}} \cdots \overline{P}^{\mu_n}
                \; [\overline{P}^{\nu)}, \Delta^{\beta }] \non \\
        + && \!\!\!\!\!\!
            \overline{U}(P') \; [\gamma^{(\mu}, \Delta^{\alpha}] \; U(P)
                \; \sum_{i=0}^{[\frac{n}{2}]}
            C_{n,2i\!-\!1} \; \Delta^{\mu_1}\Delta^{\mu_2} \cdots \Delta^{\mu_{2i\!-\!1}}
                \overline{P}^{\mu_{2i}} \cdots \overline{P}^{\mu_n}
                \; [\overline{P}^{\nu)}, \Delta^{\beta }] \non \\
        + && \!\!\!\!\!\!
            \overline{U}(P') \; [\overline{P}^{(\mu}, \Delta^{\alpha}] \; U(P)
                \; \sum_{i=0}^{[\frac{n}{2}]}
            D_{n,2i\!-\!1} \; \Delta^{\mu_1}\Delta^{\mu_2} \cdots \Delta^{\mu_{2i\!-\!1}}
                \overline{P}^{\mu_{2i}} \cdots \overline{P}^{\mu_n}
                \; [\overline{P}^{\nu)}, \Delta^{\beta }] \;. \non \\
\eeqar

\nid We have used the same symbols as for the quark form
factors.

\section{Summary and Outlook} \label{sec:summary}

By matching of quantum numbers between cross channels, we count
the number of independent form factors in the off-forward matrix
elements of the quark and gluon helicity-flip operators
((\ref{eqn:qasymop}) and (\ref{eqn:goperator})). We found that for
a rank-$n$ ($n \ge 2$) quark helicity-flip operator that number is
$2n-1$ if $n$ is even and $2n-2$ if $n$ is odd. For the rank-$n$
($n\ge 4$) gluon helicity-flip operator that number is $2n-5$ and
$2n-6$, respectively. Also, the matrix elements of these operators
have their expansion into these form factors, together with the
appropriate tensor structure and kinematic factors, in the form of
equations (\ref{eqn:qexpansion}) and (\ref{eqn:gexpansion}). The
independent form factors emerging this way can be related to the
moments of the GPDs and thus can be used to define and extract the
GPDs themselves. One can further pursue this route to cast more
light on the intrinsic structure of these operators and/or form
factors, eg, power counting (of light cone wave-functions)
\cite{ref:Ji03LC}.

As a final note, after this work was completed, a paper dealing with similar topics
emerged by P. Hagler \cite{ref:Hagler}, in which similar results for the quark operators
were obtained and were consistent with ours.

\vspace*{1cm}

\acknowledgements

Z. Chen was partly supported by a research grant from Manhattanville College.
X. Ji. was supported by the U. S. Department of Energy via grant DE-FG02-93ER-40762.
Z.C. would like to thank the U. Maryland TQHN group for their hospitality. Z.C.
would also like to thank Ying Wang, without whose encouragement and support this
work would not have been possible.

\appendix

\section{Number of Independent Elements of Tensors} \label{sec:tensor}

Using Young tableau, one can calculate the number of independent
elements of [n,1] \cite{ref:Cheng&Li} in $4$-D, and the matching
representation enumerations.
Following the discussions in \cite{ref:Hamermesh}, the systematic
decomposition of tensor
$[n,1]$ into traceless tensors 
is as follows (omitting the comma between numbers):
\beqar{eqn:qtrace}
&& [11] = (11) \non \\
&& \;\; 6 \;\;\;\;\;\;\;\;\; 6  \non \\
&& [21] = (21) + (10) \non \\
&& \; 20 \;\;\;\;\;\;\; 16 \;\;\;\;\;\;\;\; 4 \non \\
&& [31] = (31) + (11) + (20) \non \\
&& \; 45 \;\;\;\;\;\;\; 30 \;\;\;\;\;\;\;\; 6 \;\;\;\;\;\;\;\;\; 9 \non\\
&& [41] = (41) + (21) + (30) + (10) \non \\
&& \; 84 \;\;\;\;\;\;\; 48 \;\;\;\;\;\;\; 16 \;\;\;\;\;\;\;\, 16 \;\;\;\;\;\;\;\; 4 \non\\
&& [51] = (51) + (31) + (11) + (40) + (20) \non\\
&& 140   \;\;\;\;\;\;\, 70 \;\;\;\;\;\;\; 30 \;\;\;\;\;\;\;\; 6
\;\;\;\;\;\;\;\; 25
        \;\;\;\;\;\;\;\; 9 \non \\
&& \cdots \non \\
n \;\;\;\; odd  && [n,1] = (n,1)+(n\!-\!2,1) + \cdots + (11) +
(n\!-\!1,0) + (n\!-\!3,0)
                        + \cdots + (20) \non \\
n \;\; even && [n,1] = (n,1)+(n\!-\!2,1) + \cdots + (21) +
(n\!-\!1,0) + (n\!-\!3,0)
                        + \cdots + (10) \non
\eeqar

To prove the result that $(n,1) = 2n(n+2)$, one can use the
iteration method. From
\beq{eqn:n-1}
    [n,1] = (n,1)+(n\!-\!2,1) + \cdots + (n\!-\!1,0) + (n\!-\!3,0) + \cdots \;,\non
\eeq

\nid one has
\beqar{eqn:nplus2-1}
    [n\!+\!2,1] &=& (n\!+\!2,1)+(n,1) + \cdots + (n\!+\!1,0) + (n\!-\!1,0) + \cdots \non \\
                &=& (n\!+\!2,1) + (n\!+\!1,0) + [n,1] \;. \non
\eeqar

\nid Let $ D(n) = 2n(n+2)$, then it is apparent from explicit
calculation/enumeration that $D(n) = (n,1)$ is valid for
$n=1,2,3$. Given $[n,2] = \frac{n}{2}(n+2)(n+3)$ and $(n,0) =
(n+1)^2$, one has \beqar{eqn:recursion}
    [n\!+\!2,1]-[n,1]-(n\!+\!1,0) &=& \frac{n+2}{2}(n\!+\!4)(n\!+\!5)
        - \frac{n}{2}(n\!+\!2)(n\!+\!3) - (n\!+\!2)^2 \non \\
    &=& 2 (n\!+\!2)(n\!+\!4) = D(n+2) \;.\non
\eeqar

The Proof of the iteration of \beq{eqn:iteration}
    [n,1] = (n,1) + [n\!-\!2,1] + (n\!-\!1,0)
\eeq \nid is as the following:

The trace conditions for a Young tableau is the regular removal of
two boxes in the graphic representation/calculation of Young
tableaus(eg, see \cite{ref:Hamermesh, ref:Young}). And for
$[n,1]$, which is represented as

\vspace*{0.3cm}

\unitlength0.5cm
\begin{picture}(30,1)
\linethickness{0.075mm} \put(1,-1){\framebox(1,1){$\alpha$}}
\put(1,0){\framebox(1,1){$\mu_1$}}
\put(2,0){\framebox(1,1){$\mu_2$}}
\put(3,0){\framebox(3,1){$\ldots$}}
\put(6,0){\framebox(1,1){$\mu_n$}}

\put(7.8,0){=}

\put(9,-1){\framebox(1,1){$\alpha$}}
\put(9,0){\framebox(1,1){$\mu_1$}}
\put(10,0){\framebox(1,1){$\mu_2$}}
\put(11,0){\framebox(3,1){$\ldots$}}
\put(14,0){\framebox(1,1){$\mu_n$}}
\put(9,1.2){\line(1,0){6}}

\put(16,0){-}

\put(17,-1){\framebox(1,1){$\mu_1$}}
\put(17,0){\framebox(1,1){$\alpha$}}
\put(18,0){\framebox(1,1){$\mu_2$}}
\put(19,0){\framebox(3,1){$\ldots$}}
\put(22,0){\framebox(1,1){$\mu_n$} \;,}
\put(18,1.2){\line(1,0){5}}
\end{picture}

\vspace*{0.5cm}

\nid where the double line means explicitly symmetric, there are
only two ways for the contraction.

No.1. Contracting any pair out of {$\mu_2,\mu_3,...,\mu_n$}.
Because of the explicit symmetry of {$\mu_2,\mu_3,...,\mu_n$},
each contraction is the same as contracting $\mu_{n-1}$ and
$\mu_n$, This results in

\vspace*{0.3cm} \unitlength0.5cm
\begin{picture}(30,1)
\linethickness{0.075mm} \put(1,-1){\framebox(1,1){$\alpha$}}
\put(1,0){\framebox(1,1){$\mu_1$}}
\put(2,0){\framebox(1,1){$\mu_2$}}
\put(3,0){\framebox(3,1){$\ldots$}}
\put(6,0){\framebox(2,1){$\mu_{n-2}$}}

\put(8.8,0){$=$}

\put(10,0){$[n\!-\!2,1]$ \;.}

\end{picture}
\vspace*{0.6cm}

No.2. Contracting one of $1$ or $\alpha$ with any one of
{$\mu_2,\mu_3,...,\mu_n$}. Because of the anti-symmetry of $\mu_1$
and $\alpha$, we have

\vspace*{0.3cm} \unitlength0.5cm
\begin{picture}(30,1)
\linethickness{0.075mm} \put(1,-1){\framebox(1,1){$\alpha$}}
\put(1,0){\framebox(1,1){$\cdot$}}
\put(2,0){\framebox(1,1){$\cdot$}}
\put(3,0){\framebox(3,1){$\ldots$}}
\put(6,0){\framebox(1,1){$\mu_n$}}

\put(7.8,0){$= \; - \,$}

\put(10,-1){\framebox(1,1){$\cdot$}}
\put(10,0){\framebox(1,1){$\alpha$}}
\put(11,0){\framebox(1,1){$\cdot$}}
\put(12,0){\framebox(1,1){$\mu_3$}}
\put(13,0){\framebox(3,1){$\ldots$}}
\put(16,0){\framebox(1,1){$\mu_n$}}

\put(17.8,0){$= - \,$}

\put(20,-1){\framebox(1,1){$\cdot$}}
\put(20,0){\framebox(1,1){$\alpha$}}
\put(21,0){\framebox(1,1){$\mu_2$}}
\put(22,0){\framebox(3,1){$\ldots$}}
\put(25,0){\framebox(2,1){$\mu_{n-1}$}}
\put(27,0){\framebox(1,1){$\cdot$} \;.}
\end{picture}
\vspace*{0.6cm}

\nid Thus each and all of the contraction is equivalent to
contracting $\alpha$ and $\mu_n$, which results in a mixed
symmetry of {$\mu_2,\mu_3, ..., \mu_{n\!-\!1}$} (with $\mu_1$
still in front, and this is not $[n\!-\!1,0]$). Explicitly, it is,

\vspace*{0.3cm} \unitlength0.5cm
\begin{picture}(30,1)
\linethickness{0.075mm} \put(1,-1){\framebox(1,1){$\cdot$}}
\put(1,0){\framebox(1,1){$\mu_1$}}
\put(2,0){\framebox(1,1){$\mu_2$}}
\put(3,0){\framebox(3,1){$\ldots$}}
\put(6,0){\framebox(2,1){$\mu_{n\!-\!1}$}}
\put(8,0){\framebox(1,1){$\cdot$}}

\put(9.8,0){$=$}

\put(11,-1){\framebox(1,1){$\cdot$}}
\put(11,0){\framebox(1,1){$\mu_1$}}
\put(12,0){\framebox(1,1){$\mu_2$}}
\put(13,0){\framebox(3,1){$\ldots$}}
\put(16,0){\framebox(2,1){$\mu_{n\!-\!1}$}}
\put(18,0){\framebox(1,1){$\cdot$}}
\put(11,1.2){\line(11,0){7}}

\put(20,0){$-$}

\put(22,-1){\framebox(1,1){$\mu_1$}}
\put(22,0){\framebox(1,1){$\cdot$}}
\put(23,0){\framebox(1,1){$\mu_2$}}
\put(24,0){\framebox(3,1){$\ldots$}}
\put(27,0){\framebox(2,1){$\mu_{n\!-\!1}$}}
\put(29,0){\framebox(1,1){$\cdot$}}
\put(23,1.2){\line(11,0){6}}

\end{picture}
\vspace*{0.6cm}

\vspace*{0.3cm} \unitlength0.5cm
\begin{picture}(30,1)
\linethickness{0.075mm}

\put(1,0){$=$}

\put(2,-1){\framebox(1,1){$\cdot$}}
\put(2,0){\framebox(1,1){$\mu_1$}}
\put(3,0){\framebox(1,1){$\mu_2$}}
\put(4,0){\framebox(2,1){$\ldots$}}
\put(6,0){\framebox(2,1){$\mu_{n\!-\!1}$}}
\put(8,0){\framebox(1,1){$\cdot$}}
\put(2,1.2){\line(11,0){6}}

\put(9.5,0){$-\frac{1}{2}\left[ \right.$}

\put(11.5,-1){\framebox(1,1){$\mu_1$}}
\put(11.5,0){\framebox(1,1){$\cdot$}}
\put(12.5,0){\framebox(1,1){$\mu_2$}}
\put(13.5,0){\framebox(2,1){$\ldots$}}
\put(15.5,0){\framebox(2,1){$\mu_{n\!-\!1}$}}
\put(17.5,0){\framebox(1,1){$\cdot$}}
\put(12.5,1.2){\line(11,0){5}}

\put(19,0){$+$}

\put(20,-1){\framebox(1,1){$\cdot$}}
\put(20,0){\framebox(1,1){$\mu_1$}}
\put(21,0){\framebox(1,1){$\mu_2$}}
\put(22,0){\framebox(2,1){$\ldots$}}
\put(24,0){\framebox(2,1){$\mu_{n\!-\!1}$}}
\put(26,0){\framebox(1,1){$\cdot$}} \put(21,1.2){\line(11,0){5}}
\put(27.5,0){$\left. \right]$}
\end{picture}
\vspace*{0.6cm}

\vspace*{0.3cm} \unitlength0.5cm
\begin{picture}(30,1)
\linethickness{0.075mm}

\put(9.5,0){$+\frac{1}{2}\left[ \right.$}

\put(11.5,-1){\framebox(1,1){$\cdot$}}
\put(11.5,0){\framebox(1,1){$\mu_1$}}
\put(12.5,0){\framebox(1,1){$\mu_2$}}
\put(13.5,0){\framebox(2,1){$\ldots$}}
\put(15.5,0){\framebox(2,1){$\mu_{n\!-\!1}$}}
\put(17.5,0){\framebox(1,1){$\cdot$}}
\put(12.5,1.2){\line(11,0){5}}

\put(19,0){$-$}

\put(20,-1){\framebox(1,1){$\mu_1$}}
\put(20,0){\framebox(1,1){$\cdot$}}
\put(21,0){\framebox(1,1){$\mu_2$}}
\put(22,0){\framebox(2,1){$\ldots$}}
\put(24,0){\framebox(2,1){$\mu_{n\!-\!1}$}}
\put(26,0){\framebox(1,1){$\cdot$}} \put(21,1.2){\line(11,0){5}}
\put(27.5,0){$\left. \right] \, ,$}
\end{picture}
\vspace*{0.6cm}

\nid where we have rewritten the second term into a symmetric
combination of $\mu_1$ and (the contracted) $\alpha$ (the first
square parentheses) and an anti- symmetric one (the second square
parentheses).

\nid The first square parentheses represents a pair of contraction
in the now all symmetrized {$\mu_1, \mu_2, \cdots,
\mu_{n\!-\!1}$}, and together with the first term, they are
actually

\vspace*{0.5cm} \unitlength0.5cm
\begin{picture}(30,1)
\linethickness{0.075mm} \put(1,0){\framebox(1,1){$\mu_1$}}
\put(2,0){\framebox(1,1){$\mu_2$}}
\put(3,0){\framebox(3,1){$\ldots$}}
\put(6,0){\framebox(2,1){$\mu_{n-1}$}} \put(8.5,0.2){$- \;\;
\mathrm{trace~terms}$}

\put(14,0.2){$=$}

\put(15,0.2){$(n\!-\!1,0)$ \;.}

\end{picture}
\vspace*{0.6cm}

\nid On the other hand, the second square parentheses term is
symmetric in {$\mu_2, \cdots, \mu_{n\!-\!1}$} but anti-symmetric
in {$\mu_1 \leftrightarrow \mu_2$} Therefore it is indeed

\vspace*{0.3cm} \unitlength0.5cm
\begin{picture}(30,1)
\linethickness{0.075mm} \put(1,-1){\framebox(1,1){$\mu_1$}}
\put(1,0){\framebox(1,1){$\mu_2$}}
\put(2,0){\framebox(1,1){$\mu_3$}}
\put(3,0){\framebox(3,1){$\ldots$}}
\put(6,0){\framebox(2,1){$\mu_{n-1}$}}

\put(8.8,0){$=$}

\put(10,0){$[n\!-\!2,1]$ \;.}

\end{picture}
\vspace*{0.6cm}

\nid This of course is the same as No.1, ie, they are the same
trace(less) conditions.

Combine No.1 and No.2 we find that the number of trace(less)
conditions for $[n,1]$ is $[n\!-\!2,1] + (n\!-\!1,0)$, thus
proving (\ref{eqn:iteration}).

Therefore, we have the following table as the resulting
enumerations: (Also included are the numbers of
trace(less) conditions for each tensor, denoted by $Tr$, and
the column {\sl Representation} is the enumeration of the
corresponding Lorentz group representation(s) \{A,B\} (and \{B,A\}.)
\begin{table}[ht]
\vspace{6pt}
\caption{The enumeration of independent elements of tensors with $n$-symmetrized
and one pair of anti-symmetrized indices. }
\label{tab:gluon}
\begin{center}
\begin{tabular}{||c|c|c|c|c||}
\hline\hline
$ ~n~ $ & Generic & Tr & Traceless & Representation \\
\hline
$1$ & N/A & $0$ & $(1,1) = 6$ & $\{0,1\} + \{1,0\} = 3 \times 2$ \\
\hline
$2$ & $[2,1]=20$ & $4$ & $(2,1) = 16$
    & $\{\frac{1}{2},\frac{3}{2}\} \!+\! \{\frac{3}{2},\frac{1}{2}\} = 4 \times 2 $\\
\hline
$3$ & $[3,1]=45$ & $15$ & $(3,1) = 30$
    & $\{1,2\} \!+\! \{2,1\} = 15 \times 2$\\
\hline
$4$ & $[4,1]=84$ & $36$ & $(4,1) = 48$
    & $\{\frac{3}{2},\frac{5}{2}\} \!+\! \{\frac{5}{2},\frac{3}{2}\} = 24 \times 2$\\
\hline
$5$ & $[5,1]=140$ & $70$ & $(5,1) = 70$
    & $\{2,3\} \!+\! \{3,2\} = 35 \times 2$\\
\hline
$\cdots$ & $\cdots$ & $\cdots$ & $\cdots$ & $\cdots$\\
\hline
$n$ & $~[n,1] = \frac{n}{2} (n+2)(n+3)~$ & $~\frac{n}{2}(n-1)(n+2)~$
    & $~(n,1) = 2n(n+2)~$
    & $~\{\frac{n-1}{2},\frac{n+1}{2}\} \!+\! \{\frac{n+1}{2},\frac{n-1}{2}\}~$\\
& & $=C^n_2 (n+2)$ & & $= n(n+2) \times 2$\\
\hline\hline
\end{tabular}
\end{center}
\end{table}

\newpage

Similar to the case earlier, following the same
discussions in \cite{ref:Hamermesh}, the systematic decomposition
of tensor $[n+2,2]$
into traceless tensors is as follows: 
\beqar{eqn:gtrace}
&& [22] = (22) + (20) + (00) \non \\
&& \; 20 \;\;\;\;\;\;\; 10 \;\;\;\;\;\;\;\; 9 \;\;\;\;\;\;\;\;\; 1 \non\\
&& [32] = (32) + (21) + (30) + (10) \non \\
&& \; 60 \;\;\;\;\;\;\; 24 \;\;\;\;\;\;\; 16 \;\;\;\;\;\;\;\, 16 \;\;\;\;\;\;\;\; 4 \non\\
&& [42] = (42) + (40) + (20) + (31) + (22) + (20) + (00) \non\\
&& 126   \;\;\;\;\;\;\, 42 \;\;\;\;\;\;\; 25 \;\;\;\;\;\;\;\; 9
\;\;\;\;\;\;\;\; 30
        \;\;\;\;\;\;\; \underbrace{ 10 \;\;\;\;\;\;\;\; 9 \;\;\;\;\;\;\;\; 1} \non \\
&& \hspace*{7.5cm} [22] \non \\
&& [52] = (52) + (50) + (30) + (41) + (32) + (21) + (30) + (10) \non\\
&& 224   \;\;\;\;\;\;\, 64 \;\;\;\;\;\;\; 36 \;\;\;\;\;\;\, 16
\;\;\;\;\;\;\;\, 48
        \;\;\;\;\;\; \underbrace{24 \;\;\;\;\;\;\; 16 \;\;\;\;\;\;\; 16
            \;\;\;\;\;\;\;\; 4} \non \\
&& \hspace*{8cm} [32] \non \\
&& [62] = (62) + (60) + (40) + (51) + (42) + (40) + (20) + (31) + (22) + (20) + (00) \non\\
&& 360  \;\;\;\;\;\;\, 90 \;\;\;\;\;\;\; 49 \;\;\;\;\;\;\, 25
\;\;\;\;\;\;\;\, 70
        \;\;\;\;\;\;\- \underbrace{42 \;\;\;\;\;\;\; 25 \;\;\;\;\;\;\;\; 9 \;\;\;\;\;\;\;\; 30
        \;\;\;\;\;\;\; 10 \;\;\;\;\;\;\;\; 9 \;\;\;\;\;\;\;\; 1} \non \\
&& \hspace*{10cm} [42] \non \\
&& \cdots \non \\
&& [n+2,2] = (n+2,2)+(n\!+\!2,0) + (n,0) + (n\!+\!1, 1) + [n,2]
\;. \non \eeqar

\nid The proof of the iteration of $[n+2,2]$ would be similar to
before.

Again by calculating the number of elements in Young tableau, we
have $[n+2,2] = (n+1)(n+4)(n+5)$, $[n,2] = (n-1)(n+2)(n+3)$,
$(n+2,0)=(n+3)^2$, $(n,0)=(n+1)^2$, and $(n+1,1) = 2 (n+1)(n+3)$.

\section{Discussions on Time Reversal} \label{sec:Time}

Let $\hat{\mathcal{T}}$ be the time reversal operator in Hilbert
space. Then its operation on $p^\mu$ will result in $p^0
\rightarrow p^0$ while $\vec{p} \rightarrow -\vec{p}$, that is,
$\hat{\mathcal{T}} p^\mu \rightarrow p_\mu$. For spin wave
functions (of Dirac spinors) $\psi(t,\vec{x})$ one needs
$\hat{\mathcal{T}} \psi(t,\vec{x}) \hat{\mathcal{T}}^{-1}
\leftrightarrow T \psi(t,\vec{x}) \rightarrow \psi(-t, \vec{x})$,
where $T$ is a $4\times4$ time reversal matrix acting on spinors
(in Dirac space). $T$ is a so-called {\it anti-linear} or {\it
anti-unitary} operator with $T^\dag = T^{-1}$ but $T(c-number) =
(c-number)^*T$. It reverses the momentum of a particle as well as
its spin, eg, for the Fermion annihilation operators $T
b_{\vec{p}, \lambda} T^{-1} = b_{-\vec{p}, -\lambda}$ and $T
d_{\vec{p}, \lambda} T^{-1} = d_{-\vec{p}, -\lambda}$.

Labelling $\tilde{p}=(p^0, -\vec{p})$, one must have $T\ket{P} =
\eta_T \ket{\tilde{p}} \equiv e^{i\phi} \ket{\tilde{p}}$, where
$\eta_T$ is a pure phase factor. Because $\hat{\mathcal{T}}$
commutes with Lorentz boost (a change in $\vec{p}$), $\eta_T$ does
not depend on $p$ and is a fixed phase. With $\ket{p'} = e^{i
\chi} \ket{P}$, and the fact that $T$ acts on $c$-numbers is
equivalent to taking the complex conjugate, one has $T \ket{p'} =
e^{-i\chi} e^{i \phi} \ket{\tilde{p}} = e^{-2i\chi} e^{i\phi}
\ket{\tilde{p}'}$. Therefore, one can always choose $\chi =
\phi/2$ and thus
\beq{eqn:Tonp}
    T \ket{p^{(')}} = \ket{ \tilde{p}^{(')}} \;. \non
\eeq

\nid Explicitly,
\beq{eqn:psi}
    \psi(t,\vec{x}) = \sum_\lambda \int \frac{d^3 \vec{k}}{2k^0 (2\pi)^3}
        \left( e^{-ik \cdot x} b_{k,\lambda} U(k,\lambda)
            + e^{ik \cdot x} d^\dag_{k,\lambda} V(k,\lambda) \right) \non
\eeq

\nid and
\beq{eqn:TPsi1}
    \hat{\mathcal{T}} \psi(t,\vec{x}) \hat{\mathcal{T}}^{-1} =
        \sum_\lambda \int \frac{d^3 \vec{k}}{2k^0 (2\pi)^3}
            \left( e^{+ik \cdot x} b_{\tilde{k},\lambda} U^*(k,\lambda)
            + e^{-ik \cdot x} d^\dag_{\tilde{k},\lambda} V^*(k,\lambda) \right) \;. \non
\eeq

\nid Thus one needs a time reversal matrix $T$ such that
\cite{ref:IZ} $U^*(k,\lambda) = T U(\tilde{k}, \lambda)$, that is,
$TU(k) = U^*(\tilde{k})$. And one has
\beqar{eqn:TPsi2}
    \hat{\mathcal{T}} \psi(t,\vec{x}) \hat{\mathcal{T}}^{-1} &=&
        T \sum_\lambda \int \frac{d^3 \vec{k}}{2k^0 (2\pi)^3}
            \left( e^{+ik \cdot x} b_{\tilde{k},\lambda} U(\tilde{k},\lambda)
            + e^{-ik \cdot x} d^\dag_{\tilde{k},\lambda} V(\tilde{k},\lambda) \right) \non \\
        &=& T \sum_\lambda \int \frac{d^3 \vec{k}}{2k^0 (2\pi)^3}
            \left( e^{+i \tilde{k} \cdot x} b_{\tilde{k},\lambda} U(\tilde{k},\lambda)
            + e^{-i\tilde{k} \cdot x} d^\dag_{\tilde{k},\lambda} V(\tilde{k},\lambda) \right) \;, \non
\eeqar

\nid where in the last step we have changed the integration
variable from $\vec{k}$ to $-\vec{k}$. Relabel $\tilde{k} \cdot x
= k^0 x^0 + \vec{k} \cdot \vec{x} = -k \cdot \tilde{x}$, and again
change change the integration variable $\vec{k} \rightarrow
-\vec{k}$, we have
\beq{eqn:TPsi3}
    \hat{\mathcal{T}} \psi(t,\vec{x}) \hat{\mathcal{T}}^{-1} =
        T \sum_\lambda \int \frac{d^3 \vec{\tilde{k}}}{2k^0 (2\pi)^3}
            \left( e^{-i \tilde{k} \cdot \tilde{x}} b_{\tilde{k},\lambda} U(\tilde{k},\lambda)
            + e^{i\tilde{k} \cdot \tilde{x}} d^\dag_{\tilde{k},\lambda} V(\tilde{k},\lambda) \right) \;.
                \non
\eeq

\nid That is \beq{eqn:TonPsi}
    \hat{\mathcal{T}} \psi(t,\vec{x}) \hat{\mathcal{T}}^{-1}
    = T \psi(\tilde{x}) = T \psi(-t, \vec{x}) \;.
\eeq

\nid At the same time, since $\hat{\mathcal{T}}$ acts in Hilbert
space it commutes with $\gamma^0$, we have
\beqar{eqn:TonPsibar}
    \hat{\mathcal{T}} \overline{\psi}(x) \hat{\mathcal{T}}^{-1}
        &=& \hat{\mathcal{T}} \psi^\dag \gamma^0 \hat{\mathcal{T}}^{-1}
        = \hat{\mathcal{T}} \psi^\dag \hat{\mathcal{T}}^{-1} \gamma^0 \non \\
        &=& (\hat{\mathcal{T}} \psi \hat{\mathcal{T}}^{-1})^\dag \gamma^0
        = (T \psi(\tilde{x}))^\dag \gamma^0 = \psi^\dag(\tilde{x}) T^\dag \gamma^0 \non \\
        &=& \overline{\psi}(\tilde{x}) T^\dag \;.
\eeqar

As an example, take ${\cal O}^\mu = \overline{\psi} \gamma^\mu
\psi$, we have
\beqar{eqn:gamma}
    \hat{\mathcal{T}} {\cal O}^\mu \hat{\mathcal{T}}^{-1} &=&
        \hat{\mathcal{T}} \overline{\psi}(x) \hat{\mathcal{T}}^{-1}
        \hat{\mathcal{T}} \gamma^\mu \hat{\mathcal{T}}^{-1}
        \hat{\mathcal{T}} \psi(x) \hat{\mathcal{T}}^{-1} \non \\
    &=& \overline{\psi}(\tilde{x}) T^\dag (\gamma^\mu)^* T \psi(\tilde{x}) \;. \non
\eeqar

\nid In Dirac representation, $T=i \gamma^1\gamma^3$. Therefore
under $T^\dag (\gamma^\mu)^* T$, $\gamma^0 \rightarrow \gamma^0$
because $\gamma^0$ is real and commutes with $T$, $\gamma^2
\rightarrow -\gamma^2$ because $\gamma^2$ is imaginary and
commutes with $T$, and $\gamma^{1/3} \rightarrow -\gamma^{1/3}$
because $\gamma^{1/3}$ is real and anti-commutes with $T$. Thus,
since $(\gamma^i)^2 = I $ ($i = 1, 2, 3$), \beq{eqn:Tgamma}
    \hat{\mathcal{T}} \overline{\psi} \gamma^\mu \psi \hat{\mathcal{T}}^{-1}
        =  \overline{\psi} \gamma_\mu \psi \;.
\eeq

\nid Similarly, for ${\cal O}^{\mu\nu} = \overline{\psi}(x)
\sigma^{\mu\nu}\psi(x)$, \beq{eqn:Tsigma}
    \hat{\mathcal{T}} \overline{\psi} \sigma^{\mu\nu} \psi \hat{\mathcal{T}}^{-1}
        =  - \overline{\psi} \sigma_{\mu\nu} \psi \;.
\eeq

\nid And also \beqar{eqn:Tu}
    U^*(\hat{\mathcal{T}} p) &=& T U(P)  \;\;\; (or \;\;
        U(\hat{\mathcal{T}} p) = T^* U^*(p)) \non \\
    \overline{U}^*(\hat{\mathcal{T}} p') &=& \overline{U}(P') T^\dag
\eeqar

\nid For completeness, the covariant derivative $i \stackrel
{\leftrightarrow}{\cal D}$ is even ("+") under time reversal.


As an explicit example, let us now discuss the (additional)
constraints time reversal invariance might have on the matrix
elements of the lowest rank operator in (\ref{eqn:qasymop}) ${\cal
O}^{\mu\nu} = \overline{\psi}(x) \sigma^{\mu\nu}\psi(x)$. To
assure the antisymmetry, only terms of the following types
according to their Lorentz/tensorial structure will appear
\beqar{eqn:oexp}
    \bra{P'} {\cal O}^{\mu\nu} \ket{P} &=& c_1 \overline{U}(P') \sigma^{\mu\nu} U(P)
        \;\; + \;\; c_2 \overline{U}(P') [\gamma^\mu ,\overline{P}^\nu] U(P) \non \\
    && \;\; + \;\; c_3 \overline{U}(P') [\gamma^\mu ,\Delta^\nu] U(P)
        \;\; + \;\; c_4 \overline{U}(P') [\overline{P}^\mu, \Delta^\nu] U(P) \;,
\eeqar \nid where we have used the notation
$$[\gamma^\mu ,\overline{P}^\nu] = \gamma^\mu \overline{P}^\nu
                - \gamma^\nu \overline{P}^\mu$$

\nid and similar for $[\gamma^\mu ,\Delta^\nu]$ and
$[\overline{P}^\mu, \Delta^\nu]$.

The fact that ${\cal O}^{\mu\nu}$ is a Hermitian operator means
\beq{eqn:HermO}
    \bra{P'} {\cal O}^{\mu\nu} \ket{P}^\dag = \bra{p} ({\cal O}^{\mu\nu})^\dag \ket{p'}
            = \bra{p} {\cal O}^{\mu\nu} \ket{p'} \;.
\eeq

Because $\overline{P}$ is symmetric in $p$ and $p'$, and $\Delta$ is
anti-symmetric in them, the signs of $c_1$ and $c_2$ terms would
remain the same under Hermitian operation, while those of the
other two would change. That is, we have \beqar{eqn:HermOp}
    \bra{P'} {\cal O}^{\mu\nu} \ket{P}^\dag &=& \bra{p} {\cal O}^{\mu\nu}
        \ket{p'} \non \\
    &=& \;\; c_1 \overline{U}(P) \sigma^{\mu\nu} U(P')
        \;\; + \;\; c_2 \overline{U}(P) [\gamma^\mu ,\overline{P}^\nu] U(P') \non \\
    && \;\; - \;\; c_3 \overline{U}(P) [\gamma^\mu ,\Delta^\nu] U(P')
        \;\; - \;\; c_4 \overline{U}(P) [\overline{P}^\mu, \Delta^\nu] U(P') \;.
\eeqar

For the $c_1$ term, direct application of the Hermitian would
give, \beqar{eqn:c1}
    \bra{P'} {\cal O}^{\mu\nu} \ket{P}^\dag &=&
            \cdots + c_1^* [\overline{U}(P') \sigma^{\mu\nu} U(P)]^\dag + \cdots
            = \cdots + c_1^* U(P)^\dag (\sigma^{\mu\nu})^\dag \overline{U}(P')^\dag
                + \cdots \non \\
            &=& \cdots + c_1^* U(P)^\dag \gamma^0 \sigma^{\mu\nu} \gamma^0 \gamma^0 U(P')
               +\cdots = \cdots + c_1^* \overline{U}(P) \sigma^{\mu\nu} U(P')
               + \cdots ;.
\eeqar

\nid Therefore one has $c_1^*=c_1$ which means $c_1$ is real, and
let us label $c_1 \equiv  C_1$.

\nid Similarly, for the $c_2$ term, direct Hermitian gives
\beqar{eqn:c2}
    \bra{P'} {\cal O}^{\mu\nu} \ket{P}^\dag &=&
            \cdots + c_2^* U(P)^\dag (\gamma^\mu \overline{P}^\nu
                - \gamma^\nu \overline{P}^\mu)^\dag \overline{U}(P')^\dag
                + \cdots \non \\
            &=& \cdots + c_2^* U(P)^\dag (\gamma^0\gamma^\mu\gamma^0 \overline{P}^\nu
                - \gamma^0\gamma^\nu\gamma^0 \overline{P}^\mu) \gamma^0 U(P')
                + \cdots \non\\
            &=& \cdots + c_2^*\overline{U}(P) [\gamma^\mu ,\overline{P}^\nu] U(P') \;.
\eeqar

\nid Therefore one has $c_2^*=c_2$ which means $c_2$ is also real,
and let us label $c_2 \equiv  C_2$.

\nid For the $c_3$ term, on the other hand, direct Hermitian gives
\beqar{eqn:c3}
    \bra{P'} {\cal O}^{\mu\nu} \ket{P}^\dag &=&
            \cdots + c_3^* U(P)^\dag [\gamma^\mu ,\Delta^\nu]^\dag \overline{U}(P')^\dag
            + \cdots \non \\
        &=& \cdots + c_3^* U(P)^\dag \gamma^0 [\gamma^\mu ,\Delta^\nu] \gamma^0 \gamma^0 U(P')
            + \cdots \non\\
        &=& \cdots + c_3^*\overline{U}(P)[\gamma^\mu ,\Delta^\nu] U(P') \;.
\eeqar

\nid Compared with (\ref{eqn:HermOp}) one has $c_3^*=-c_3$ which
means $c_3$ is pure imaginary, and one defines $c_3 \equiv i C_3$
where $C_3$ is real.

\nid And for the $c_4$ term, \beqar{eqn:c4}
    \bra{P'} {\cal O}^{\mu\nu} \ket{P}^\dag
        &=& \cdots + c_4^* U(P)^\dag [\overline{P}^\mu ,\Delta^\nu]^\dag \overline{U}(P')^\dag
            + \cdots \non \\
        &=& \cdots + c_4^* U(P)^\dag \gamma^0 [\overline{P}^\mu ,\Delta^\nu] \gamma^0 \gamma^0 U(P')
            + \cdots \non \\
        &=& \cdots + c_4^*\overline{U}(P)[\overline{P}^\mu ,\Delta^\nu] U(P') \;.
\eeqar

\nid Therefore one has $c_4^*=-c_4$ which means $c_4$ is also pure
imaginary, and one defines $c_4 \equiv i C_4$ where $C_4$ is real.

Thus, the matrix element actually has the form \beqar{eqn:OpHermf}
    \bra{P'} {\cal O}^{\mu\nu} \ket{P} &=& C_1 \overline{U}(P') \sigma^{\mu\nu} U(P)
        \;\; + \;\; C_2 \overline{U}(P') [\gamma^\mu ,\overline{P}^\nu] U(P) \non \\
    &+& \;\; i C_3 \overline{U}(P') [\gamma^\mu ,\Delta^\nu] U(P)
        \;\; + \;\; i C_4 \overline{U}(P') [\overline{P}^\mu, \Delta^\nu] U(P) \;,
\eeqar

\nid where all the $C_i (i = 1,2,3,4)$ are real.

\vspace*{1cm}

On the other hand, the requirement of time reversal invariance
means \beq{eqn:TimeO}
    \bra{P'} {\cal O}^{\mu\nu} \ket{P} = \bra{P'} T^{-1}T {\cal O}^{\mu\nu}
            T^{-1}T \ket{P}^*
            = \bra{TP'} T {\cal O}^{\mu\nu} T^{-1} \ket{TP}^* \non
\eeq

\nid The $C_1$ term in $\bra{TP'} T {\cal O}^{\mu\nu} T^{-1}
\ket{TP}^*$ is (see (\ref{eqn:Tsigma})) \beqar{eqn:Tc1}
    - (C_1^* \overline{U}(TP') \sigma_{\mu\nu} U(TP))^*
        &=& -C_1 \overline{U}^*(TP') \sigma_{\mu\nu}^* U^*(TP) \non \\
        &=& -C_1 \overline{U}(P') T^\dag \sigma_{\mu\nu}^*T U(P) \;,
\eeqar

\nid where we have used (\ref{eqn:Tu}) in the last step. Since
$T^\dag = T = i \gamma^1 \gamma^3$, we have, similar to
(\ref{eqn:Tgamma}), \beqs
    T^\dag \sigma_{\mu\nu}^*T &=& i\gamma^1\gamma^3 (-\frac{i}{2})
        [\gamma_\mu, \gamma_\nu]^* i\gamma^1\gamma^3 \non \\
    &=& \frac{i}{2}(\gamma^1\gamma^3 \gamma_\mu^*\gamma_\nu^* \gamma^1\gamma^3
        - \gamma^1\gamma^3 \gamma_\nu^*\gamma_\mu^* \gamma^1\gamma^3) \non \\
    &=& -\sigma^{\mu\nu} .
\eeqs

\nid It is clear that time reversal invariance does not impose
further constraints on this term, while the structure
$\sigma^{\mu\nu}$ is odd ("-") under time reversal.

Because the time reversal takes $C$-numbers to its complex
conjugate, together with (\ref{eqn:Tgamma}), we have under time
reversal $[\gamma^\mu ,\overline{P}^\nu] \rightarrow [\gamma_\mu
,\overline{P}^\nu]$, and the $C_2$ term becomes \beqar{eqn:Tc2}
    \bra{TP'} T C_2[\gamma^\mu ,\overline{P}^\nu] T^{-1} \ket{TP}^*
        &=& (C_2^* \overline{U}(TP') [\gamma_\mu ,\overline{P}^\nu] U(TP))^* \non \\
        &=& C_2 \overline{U}^*(TP') [\gamma_\mu ,\overline{P}^\nu]^* U^*(TP) \non \\
        &=& C_2 \overline{U}(P') T^\dag (\gamma_\mu^* \overline{P}^\nu
                - \gamma_\nu^* \overline{P}^\mu)T U(P) \;.
\eeqar

\nid Similarly, we would have
$$T^\dag (\gamma_\mu^* \overline{P}^\nu - \gamma_\nu^* \overline{P}^\mu)T
    = [\gamma^\mu, \overline{P}^\nu]$$ \;.

\nid Thus we need $C_2 = -C_2$ and hence we must have $C_2 \equiv
0$.

The $C_3$ term, on the other hand, is \beqar{eqn:Tc3}
    \bra{TP'} T iC_3[\gamma^\mu ,\Delta^\nu] T^{-1} \ket{TP}^*
        &=& (-iC_3^* \overline{U}(TP') [\gamma_\mu ,\Delta^\nu] U(TP))^* \non \\
        &=& i C_3 \overline{U}^*(TP') [\gamma_\mu ,\Delta^\nu]^* U^*(TP) \non \\
        &=& i C_3 \overline{U}(P') T^\dag (\gamma_\mu^* \Delta^\nu
                - \gamma_\nu^* \Delta^\mu)T U(P) \;,
\eeqar

\nid with
$$T^\dag (\gamma_\mu^* \Delta^\nu - \gamma_\nu^* \Delta^\mu)T
    = (\gamma^\mu \Delta^\nu - \gamma^\nu \Delta^\mu) = [\gamma^\mu, \Delta^\nu] \;.$$

\nid It is clear then that time reversal invariance does not
impose further constraints on this term.

Similarly, the $C_4$ term is \beqar{eqn:Tc4}
    \bra{TP'} T iC_4[\overline{P}^\mu ,\Delta^\nu] T^{-1} \ket{TP}^*
        &=& (-iC_4^* \overline{U}(TP') [\overline{P}^\mu ,\Delta^\nu] U(TP))^* \non \\
        &=& i C_4 \overline{U}^*(TP') [\overline{P}^\mu ,\Delta^\nu]^* U^*(TP) \non \\
        &=& i C_4 \overline{U}(P') T^\dag [\overline{P}^\mu ,\Delta^\nu] T U(P) \;,
\eeqar

\nid and obviously
$$T^\dag [\overline{P}^\mu ,\Delta^\nu] T = [\overline{P}^\mu ,\Delta^\nu] \;.$$

\nid This means time reversal invariance does not impose further
constraints on this term either.

Therefore we confirmed there are three independent form factors
for the operator ${\cal O}_{qT}^{\mu \nu}$, consistent with the
general counting in section \ref{sec:qcount}. Similar discussions
can be carried out for other helicity-flip operators, and from the
combination of Hermiticity and time reversal symmetry, the general
counting result will all stand.


\end{document}